\documentclass[sigconf]{acmart}
\AtBeginDocument{%
  \providecommand\BibTeX{{%
    \normalfont B\kern-0.5em{\scshape i\kern-0.25em b}\kern-0.8em\TeX}}}

\setcopyright{acmcopyright}
\copyrightyear{2024}
\acmYear{2024}
\acmDOI{XXXXXXX.XXXXXXX}

\acmConference[CHI '24]{Make sure to enter the correct
  conference title from your rights confirmation emai}{May 11--16,
  2024}{Woodstock, NY}
\acmPrice{15.00}
\acmISBN{978-1-4503-XXXX-X/18/06}

\begin{document}

\newcommand{\ankolika}[1]{{\small\color{blue}{\bf [ankolika: #1]}}}

\title[Situated Infrastructuring]{A Situated-Infrastructuring of WhatsApp for Business in India }


\author{Ankolika De}
\email{apd5873@psu.edu}
\orcid{}
\affiliation{%
  \institution{College of Information Sciences and Technology, The Pennsylvania State University}
  \country{USA}
}

\renewcommand{\shortauthors}{De}

\begin{abstract}
WhatsApp has become a pivotal communication tool in India, transcending cultural boundaries and deeply integrating into the nation's digital landscape. Meta's introduction of WhatsApp for Business aligns seamlessly with the platform's popularity, offering businesses a crucial tool. However, the monetization plans pose challenges, particularly for smaller businesses, in balancing revenue goals with accessibility. This study, employing discourse analysis, examines Meta's infrastructuring of WhatsApp in India, emphasizing the dynamic interplay of technological, social, and cultural dimensions. Consequently, it highlights potential power differences caused by the deployment of WhatsApp for Business followed by its gradual but significant modifications, encouraging scholars to investigate the implications and ethics of rapid technological changes, particularly for marginalized users.

\end{abstract}




\maketitle

\section{Introduction}

WhatsApp is a pivotal communication tool in India, widely embraced across diverse social and economic segments. It serves as a pervasive platform facilitating both personal and professional interactions, transcending linguistic and cultural boundaries \cite{mehta2017exploratory}. The application's user-friendly interface and accessibility contribute to its extensive usage, fostering seamless communication \cite{chanda2018whatsapp}. Its adaptability makes it a crucial asset for businesses, aiding in marketing, customer service, and collaboration \cite{bagdare2021whatsapp}. Given WhatsApp's already pervasive presence as a widely embraced communication platform, Meta's strategic move to introduce WhatsApp for Business in India aligns seamlessly with its logical evolution. Capitalizing on WhatsApp's extensive user base, this business-oriented iteration harnesses the platform's established popularity, making it an intuitive choice for businesses to leverage for their communication needs \cite{business-standard-meta-whatsapp}. WhatsApp's pivotal role in India's digital infrastructure goes beyond mere connectivity; it embeds itself deeply into the cultural fabric of the nation \cite{doi:10.1177/01968599221095177}. Leveraging India's widespread access to affordable data \cite{bbc-mobile-data-india} and advanced digital technologies, WhatsApp has seamlessly integrated into the daily lives of its culturally diverse population. The platform's simplicity and versatility make it a preferred mode of communication across various linguistic and cultural boundaries, reflecting its adaptability to India's rich tapestry of traditions. In essence, WhatsApp not only serves as a communication tool but also becomes a conduit for preserving and transmitting cultural nuances, reinforcing its importance in the intricate interplay of technology and culture within India's digital landscape \cite{doi:10.1177/01968599221095177}. 

The significance of WhatsApp for businesses, particularly small ones, is paramount in India's digital landscape. With small businesses forming the backbone of the economy \cite{chakrabarty2011micro}, their ability to leverage WhatsApp for marketing and communication is crucial for their survival and growth. WhatsApp for Business provided them with a cost-effective and efficient means to connect with customers, offer services, and collaborate. Given India's unique demographics and economics, characterized by a substantial youth population showing interest in entrepreneurship, and considering the country's technological capacities and widespread technological adoption \cite{chaudhary2017demographic, javalgi2016aspirations}, businesses swiftly integrated WhatsApp into various aspects of their operational processes. However, now, as Meta endeavors to monetize on this platform \cite{vanian2023meta}, businesses would be presented with two options: 1) persist with the existing system and fulfill the associated financial obligations, a challenge compounded by WhatsApp's historical appeal of being generally free, cost-effective, and user-friendly, or 2) explore alternative solutions, necessitating a fundamental restructuring of their business pipelines to accommodate novel tools, thereby demanding additional resources, time, and effort. This not only impacts the business landscape but also influences the broader economic dynamics of the nation, making it imperative to carefully navigate the platform's evolution to ensure its continued utility for businesses across the spectrum. Thus, I ask: \textit{\textbf{How does Meta introduce, embed and monetize WhatsApp for Business within India's business ecosystem, and how do the gradual changes in their pricing and other monetization methods affect businesses?}}

I employ discourse analysis \cite{gill2000discourse}, drawing insights from major Indian newspaper articles (n=38), Meta's official blog (n=9), and success stories on Meta's platform (n=25) to provide a triangulated perspective on WhatsApp for Business strategies in the Indian context, encompassing external perspectives, company self-presentation, and practical outcomes. Particularly guiding my work, is Star's conceptualization of infrastructure as not just built via physical systems, but rather a manifestation of sociotechnical arrangements \cite{star2002infrastructure}. I report Meta's situated infrastructuring of WhatsApp for Business in India, as characterized by the dynamic interplay of technological, social, and cultural dimensions. It leverages people's existing comfort with using WhatsApp, deeply ingrained in specific cultural and ritualistic practices, to build a stronger-situated infrastructure for business communication. Secondly, I argue that the introduction and monetization of WhatsApp for Business result in an amplification of power for already dominant entities, particularly large businesses. Simultaneously, individuals in marginalized sectors experience further disenfranchisement.

\section{Technological Innovation and Infrastructure}
Innovation and infrastructure are intertwined, with technological advancements often entailing the extension or addition of infrastructural elements. Modern infrastructure projects extend beyond physical constructions, incorporating social, environmental, and other factors, necessitating explicit connections to development plans \cite{ridley2006infrastructure}. Both infrastructure and technological innovation significantly influence economic and industrial growth in emerging economies, making them pivotal considerations for policy and technological objectives \cite{fan2018technological}.

Traditional methods for introducing new infrastructures often adopt inflexible and mechanistic approaches, assuming that all system requirements are well-defined and determined by prior tests and formal processes \cite{10.1145/192844.193021}. However, contemporary infrastructure is characterized by dynamism, contextuality, and interrelation with broader sociopolitical and cultural contexts \cite{blok2016infrastructuring}. Consequently, infrastructures have become more intricate and inseparable from their use, practice, and enactment, while appearing more elusive and challenging to comprehend \cite{bowker2010toward, 10.1145/192844.193021, harvey2016enchantments}.

Star (1999) outlined a comprehensive set of characteristics defining infrastructure, emphasizing its heightened visibility during breakdowns \cite{doi:10.1177/00027649921955326}. Implementing novel infrastructures to enhance task efficiency often relegates the infrastructure to a latent state, surfacing prominently only in instances of substantial issues during utilization. Within the context of Meta's introduction of WhatsApp for Business in India, the assimilation of this new infrastructure may appear seamless. However, potential failures become apparent when operational modifications are instituted. Star argued that changes in information infrastructures inherently influence culture and norms \cite{doi:10.1177/00027649921955326}. Such changes and implementations are often embedded in situated power relations and networks that favor some over others \cite{doi:10.1177/0042098017720149}. Graham and Marvin (2001, p.11) explicated that infrastructures upheld \textit{"sociotechnical geometries of power"} \cite{graham2001splintering}. Their conceptualization suggests that infrastructures play a pivotal role in shaping and maintaining power relations within society, forming a dynamic interplay between technological structures and sociopolitical forces that collectively define the intricate fabric of power dynamics in a given context. Scholars have underscored the importance of complementary policies alongside infrastructure investments, emphasizing that infrastructure can create opportunities but should not disproportionately benefit those already in advantageous positions \cite{bajar2015impact}. 

Bowker and Star's seminal work \cite{bowker1999sorting} defines crucial characteristics of infrastructures, emphasizing their embedded nature within diverse structures, and social arrangements. The complexity of information infrastructure is articulated, portraying it as large, layered, and multifaceted, serving diverse purposes that are contextually contingent. The process of transforming infrastructures is a time-intensive endeavor that necessitates careful negotiations and adjustments with various components within the systems under consideration \cite{10.1145/2675133.2675232}. In HCI literature, the study of information infrastructures is especially embraced in collaborative scientific projects happening across different places. Research typically focuses on understanding scientific information infrastructures, via systems bringing together people, information, and technologies \cite{10.1007/s10606-010-9114-y}. Furthermore, scholars have also been interested comprehending the role of embeddedness in infrastructure development \cite{10.1007/s10606-010-9114-y} and have confronted challenges related to expanding an infrastructure to cater to potential new users \cite{10.1145/1316624.1316660}, among other considerations. Finally, prior research has also investigated ways in which material and physical artifacts and their relationships in an infrastructural network, impact the infrastructure and vice versa \cite{10.1145/3313831.3376201}. Scholars have also directed their attention towards the concept of \textit{"infrastructuring,"} viewing it as a continuous entrepreneurial endeavor aimed at safeguarding users within specific infrastructures- in this way it is the dynamic and open-ended process of creating and evolving information infrastructures \cite{karasti2018studying}. This perspective implies a potential emphasis on individual responsibility, which is noteworthy given the intricate dependencies that exist within infrastructures \cite{10.1145/3025453.3025959}.  Drawing on Star's seminal work on infrastructure \cite{doi:10.1177/00027649921955326} and considering WhatsApp's ritualistic role in India's communication landscape \cite{doi:10.1177/01968599221095177}, the subsequent paragraph will discuss WhatsApp, its presence in India, and provide a brief overview of its business counterpart.

\section{WhatsApp in India}

WhatsApp made its debut in the Indian market in mid-2010s, transforming from a platform initially designed for sharing statuses to a comprehensive messaging application featuring group functionality for multiple users \cite{madanapalle2017whatsapp}. The rapid success of WhatsApp in India can be attributed to its role as a cost-effective substitute for conventional instant messaging services, which incurred charges per message. Moreover, the affordability and widespread accessibility of the internet further contributed to its popularity \cite{bbc-mobile-data-india}. In 2019, India secured its position as WhatsApp's largest market, with an impressive user base of around 400 million individuals \cite{iqbal2020whatsapp}. WhatsApp, extending beyond its conventional role as a communication tool, intricately embedded itself within the social and cultural fabric of India, becoming an integral component of the country's sociotechnical milieu \cite{doi:10.1177/01968599221095177}. The widespread adoption of cell phones and smartphones in India significantly contributed to the pervasive influence of WhatsApp \cite{agrawal2018india}. This phenomenon is observed across diverse age groups, where the application is employed not only for essential communication but also actively engages users in socially ingrained rituals, exemplified by routine greetings and the circulation of forwarded messages \cite{doi:10.1177/01968599221095177, mukherjee2019imagining}. Despite facing criticism for its role in disseminating misinformation, especially during the COVID-19 pandemic \cite{10.1145/3491102.3517588}, WhatsApp remained a prominent platform for information exchange across diverse geographical regions and social strata. However, some individuals opted to discontinue using WhatsApp as a reliable source of information, while retaining its usage as a social application \cite{10.1145/3512964}.

As India emerged as Meta's largest user base for WhatsApp, the challenge of monetizing the application prompted Meta's leadership to shift focus towards capturing the business marketing sector in the country \cite{kanchan2023metaturbulence}. Recognizing the prevalent use of messaging apps by businesses for promotion, sales, and customer engagement \cite{sinch2023}, Meta strategically introduced WhatsApp for Business as a dedicated pipeline for businesses. This move aimed to facilitate businesses of various sizes in performing a diverse array of tasks such as automating personalized responses to customers and labelling conversations to organize them among others \cite{businesstoday2023whatsapp}. By implementing WhatsApp for Business, Meta successfully translated India's familiarity with WhatsApp into tangible revenue.

\section{Methods}

The data collection involved systematically searching and selecting articles published between July 29, 2023 to October 29, 2023 using relevant keywords ("WhatsApp for Business", "Meta's WhatsApp", "Business on WhatsApp", "Business on WhatsApp") on Google News to ensure a diverse range of viewpoints. Primary sources comprised major national newspapers (The Financial Times, Hindustan Times, The Times of India, Business Today, and The Economic Times) (n=38), chosen for insights into emerging trends related to Meta's WhatsApp for Business. Meta's official blog posts about WhatApp for Business (n= 9) was consulted to understand the company's messaging and own utilization on WhatsApp for Business. Finally, analyzing success stories that were published on Meta's blogs (n= 25) provided a nuanced perspective on the practical implementation and impact of WhatsApp for Business in the Indian business landscape, as positioned by Meta. This triangulated approach aimed for a thorough exploration, as the newspaper articles provided an external discourse, Meta's official blog posts provided technical and corporate functioning and the success stories on Meta's blogs manifested Meta's own positioning of the technology \cite{doi:10.1177/1558689812437100}.

I used discourse analysis (DA) \cite{rapley2007doing} to analyze the datasets, triangulating insights from newspapers, Meta's blog, and success stories, to interpret the platform's strategic positioning, public perception, and practical implications, aiding in identifying potential gaps between external perceptions of the system and Meta's intended narrative. Traditionally DA focuses on \textit{how language is employed in specific contexts} \cite{rapley2007doing}- explicating how language is not neutral and deeply representative of power dynamics and sociocultural phenomena. Drawing from Jupp (2006), I annotated sentences with single codes, and further analyzed them for broader contextualization within my research question \cite{jupp2006}. Next, I separately clustered codes together to answer the research questions via introduction, embedding, and positioning of WhatsApp for Business. The newspaper articles majorly contributed to situating the introduction, while codes from Meta's blog and success stories positioned Meta's discourse during the embedding process.

\section{Findings}
In this section, I report the findings in three parts. First, I establish the dynamic in which WhatsApp for Business entered Indian markets. Next, I report how WhatsApp introduced changes in their pricing model. Finally, I discuss Meta's positioning of WhatsApp in India. 

\subsection{The Entrance of WhatsApp for Business in India}

Meta forged a partnership with the "Confederation of All India Traders (CAIT)" for their \textit{"WhatsApp Se Vyapaar"} (WhatsApp for Business) initiative (ID01)- 

\begin{quote}
    \textit{40,000 trade associations and 80 million traders across India [were trained] ... to provide digital and skill training to equip businesses with essential knowledge to help digitize their storefronts and build their ‘digital dukaan’ on the WhatsApp Business App. (ID01)}
\end{quote}

Particular insights gleaned from this initiative included the imparting of knowledge on \textit{"features like Catalog, Quick Replies and Click to WhatsApp Ads (ID02)."} Additionally, Meta disclosed its intention to \textit{"launch Meta Small Business Academy"}, a certification program geared towards empowering workers to successfully digitize their businesses.

Investing substantially in upskilling workers for effective WhatsApp utilization, Meta introduced new features that aimed to expand communication paradigms by integrating different channels. Messages on WhatsApp could now be translated to those on Facebook and Instagram (ID03). Moreover, Meta introduced features positioning WhatsApp as a comprehensive solution for businesses. This involved the creation of richer in-chat experiences through flexible, pre-made building blocks for activities such as booking appointments and placing orders. The introduction of WhatsApp Payment-to-Merchant further facilitated transactions, allowing users to add items to a cart and make payments using their preferred mode (ID03). These advancements were reinforced with pragmatic ideas and implications for the tool,

\begin{quote}
    \textit{So, for example, a bank can build a way for customers to book an appointment to open a new account, a food delivery service can build a way to place any order from any of their partner restaurants or an airline can build a way to check in for a flight and pick up a seat, all without having to leave the chat thread. (ID03)} 
\end{quote}

The integration of WhatsApp for Business was carefully planned as an extension to the original platform of simple communication \cite{mehta2017exploratory}, and transpired against the contextual backdrop of Meta's CEO, Mark Zuckerberg, claiming, \textit{"We're bringing the focus to how we can support businesses by creating simple to use and easy to scale tools."} Further elucidating the corporate vision, Sandhya Devanathan, Vice President of Meta, India, underscored India's affinity towards WhatsApp, articulating, \textit{"WhatsApp is their business — it’s their website, their digital storefront, their livelihood" (ID04)}- highlighting the all-encompassing role that WhatsApp for Business was intending to take in allowing businesses to entirely shift their digital needs on it, while still keep it a simple communication tool. Statements highlighting a pronounced focus on India may seem benevolent, ostensibly backing businesses; however, they obscure the rapid acquisition of digital commerce and communication infrastructures with a large user base by a single corporation.

The commendation for WhatsApp for Business within the Indian community was bolstered through a strategic alliance with Jio, a conglomerate with pervasive presence across diverse sectors of Indian business (ID05). This collaboration gave rise to \textit{Jiomart}, an online grocery shopping service facilitated through WhatsApp, enabling customers to place orders conveniently through the application. Additionally, another collaborative effort with PayU, an Indian payments company, was established to integrate payment functionalities within WhatsApp (ID06).

In conclusion, Meta's multifaceted approach to integrating WhatsApp for Business into India's existing infrastructure showcased a strategic effort to become a dominant cultural and commercial force in the country. In its introduction, Meta forged localized, objective, and large-scale connections, successfully demonstrating its commitment to supporting the Indian markets and building a network with key actors in India's business ecosystem. The company's initiatives extended beyond infrastructure development, aiming to transition WhatsApp into a pivotal hub for digital interactions, transactions, and business operations, capitalizing on users' integration of WhatsApp as a key communication tool to yield tangible revenue streams.

\subsection{Changes to WhatsApp's Model}

The Financial Times, in an article titled \textit{"India has embraced WhatsApp but Meta now needs to make it pay," } criticized Meta for using India as a \textit{"Petri Dish"} for WhatsApp for Business. On June 1, 2023, Meta implemented a modification to the business model of WhatsApp, transitioning from a pricing structure based on per message to per conversation. This alteration significantly impacted small businesses, as the revised business model led to escalated costs, particularly for \textit{"customer acquisition and retention" (ID07)}. The adjusted pricing framework prompted brands, particularly those with constrained capital, to exercise caution in their utilization of WhatsApp (ID07). In tandem with these alterations, Meta introduced a verification mechanism to authenticate businesses on WhatsApp, offering it as a premium subscription for those verified on Meta (ID09). Consequently, small businesses were compelled to \textit{"explore alternative platforms"} as the revised pricing policies prompted a cautious assessment of the impact of WhatsApp usage (ID09). Simultaneously, this pricing model adjustment ushered in novel features, allowing users to generate advertisements on other Meta-owned social media platforms, including Facebook and Instagram, without necessitating an account (ID08); however, it requires more technical acumen and digital agility.
 
These changes prompted concerns, with newspapers in India characterizing Meta's strategy as a retrospective plan that could have enduring negative ramifications for businesses. However, despite concerns about costs, many business owners persisted in utilizing the platform due to its integral and embedded role in their operations as their business pipelines were significantly dependant on it (ID11). 

Meta continued to introduce distinctive alterations that not only enhanced revenue but also facilitated streamlined and professionalized business operations, albeit with apparently minimal effort but substantial cost. For instance, they introduced features enabling automated personalized messages, holiday greetings, and appointment reminders (ID12), while also collaborating with rival digital payment providers, such as Razorpay and PayU, to enhance payment flexibility by allowing users to add various forms of payment methods, including credit cards, popular UPI applications, and debit cards (ID14).

Essentially, Meta initially empowered Indians to shape how they utilized WhatsApp for innovative purposes, subsequently capitalizing on these adaptations for revenue generation. This is evident from the sudden changes that are rapidly deployed, with little to no consideration for the majority of businesses that Meta had originally claimed to cater to. This raises considerations about how businesses, having integrated WhatsApp into their operations based on its initially situated factors, will adapt to or resist these significant alterations in the platform's infrastructure. Moreover, it brings in questions about the disparity in adaptability among different businesses.

\subsection{Meta's Positioning of WhatsApp in India}

Meta communicates a vision of WhatsApp as an indispensable tool for India. The strategic positioning aligns with the idea of infrastructuring by integrating WhatsApp into the fabric of daily operations and emphasizing its impact on business growth and communication paradigms, particularly reinforcing it as a testament that reflects India's embrace of new technologies. The messaging platform is portrayed as more than just a communication tool; it is positioned as a fundamental element that businesses can't afford to overlook. While launching new features for WhatsApp for Business, Mark Zuckerberg claimed, 

\begin{quote}
    \textit{India (is) a country that's at the forefront of a lot of what we're going to talk about today. You're (India is) leading the world in terms of how people and businesses have embraced messaging as the better way to get things done,} (ID14). 
\end{quote}

Preceding the aforementioned updates in 2023, anticipations had circulated in 2022 regarding the transformation of the widely embraced messaging application into a platform encompassing commercial activities. An envisioned feature akin to a \textit{yellow pages} was contemplated, where participating businesses would be introduced to other familiar establishments in their vicinity, facilitating the exchange of contact information when mutually agreed upon (ID15). The narrative implies that WhatsApp's features seamlessly integrate into the operations of businesses, creating an infrastructure where direct and efficient communication is not just an option but a necessity.  India, being a primary testing ground for Meta, not only due to its extensive user familiarity but also owing to the innovative utilization by a substantial user base, has consistently played a pivotal role in affirming the autonomous and efficient utilization of WhatsApp for both personal and business purposes. The platform's ubiquity in India was exemplified by the notable usage of channels by the country's Prime Minister, Narendra Modi, who employs them as a means to connect with the public. Through his personal channel, he communicated:

\begin{quote}
    \textit{Thrilled to join the WhatsApp community! It is yet another step closer in our journey of continued interactions. Let's stay connected here! Here’s a picture from the new Parliament Building…} (ID16)
\end{quote}

The strategic positioning and utilization of WhatsApp exhibit a distinctiveness incomprehensible to other social media platforms, offering functionalities encompassing photo sharing, business engagement, and basic messaging. The official WhatsApp website systematically disseminated a compilation of narratives elucidating the diverse attributes of businesses, irrespective of their magnitude, following the integration of WhatsApp into their operational frameworks. Meta strategically emphasized numerical metrics within these narratives, optimizing quantitative outputs that highlight the accomplishments realized by businesses through WhatsApp utilization. Accompanied by visual depictions and textual descriptions, these narratives underscored the accessibility and utility inherent in leveraging a communication platform like WhatsApp (ID(s) ss1 to ss9). For example, the utilization of the WhatsApp Business Platform is exemplified through the case of an Indian property management company as a testament. Meta wrote that the founders, initially facing challenges in the corporate sector, witnessed a remarkable 5\% increase in income after adopting WhatsApp. The platform's auto message feature, designed to handle high volumes of customer queries, played a pivotal role, contributing to the company's efficient operations. Meta underscores the transformative impact by emphasizing not just the overall growth of the business but also the significant scale, with approximately 400 properties managed across multiple locations. This narrative positions WhatsApp in India as an indispensable tool for businesses, emphasizing its integration into daily operations, its transformative impact on communication paradigms, and its ubiquitous presence in the nation's digital landscape, thereby highlighting its pivotal role in driving innovation and connectivity across the country.

\section{Discussions}

WhatsApp for Business in India unfolds through a dynamic interplay of technological, social, and cultural dimensions. Meta's active involvement in strategic partnerships and skill-building initiatives contributes to the co-construction of socio-technical infrastructures, extending beyond the digital realm. Notably, the collaboration with CAIT and the Meta Small Business Academy integrates digital and skill training into the existing business practices and socio-cultural rituals of Indian traders. This integration serves as a foundation for a socio-technical infrastructure, extending the ritualized nature of WhatsApp in India \cite{doi:10.1177/01968599221095177}. In this context, Meta strategically positions its business application for users who already appreciate, utilize, and engage with WhatsApp \cite{balkrishan2016making}. This shows that the situated enactment of WhatsApp for Business implemented within established practices \cite{balkrishan2016making} strengthens its presence in India's socio-cultural landscape, potentially increasing dependency on the platform. People through their personal experiences may shape the way WhatsApp is used for professional services --- further embedding this digital ecosystem within the daily rituals of Indian users. If access to the application becomes difficult it may pose difficulties for users to adapt or cope.  Here, I argue that the prior ritualized use of WhatsApp for communication signifies and strengthens the success of WhatsApp for Business, and has the potential to the same extent to exacerbate disparities amidst businesses if modified incessantly. 

WhatsApp, which is already deeply ingrained in Indian society, undergoes a gradual transition into more tangible forms of professional communication. Notably, Meta not only establishes a situated infrastructure by legitimizing WhatsApp from the personal to the professional but also solidifies WhatsApp as a boundary object; adaptable across multiple needs while maintaining distinct identities(\cite{star1989institutional}, p. 37). In India, where personal relationships and cultural dynamics are crucial in a successful business \cite{harriss2002trust}, WhatsApp emerges as the pragmatic choice as its users are motivated by their own situated uses to easily pursue professional communication within personal circles also. It seamlessly integrates the personal and professional aligning with cultural norms, and offering a reliable identity that fosters trust \cite{doi:10.1177/2050157920970582, harriss2002trust}. As technology, exemplified by WhatsApp for Business, becomes intricately interwoven with cultural practices, it can transcend its utilitarian functionality \cite{10.1145/2800835.2800965}. Instead, it may assume a more profound role within the societal framework, exerting mutual influence on norms, values, and rituals \cite{faik2020information, herdin2008culture}. 

However, not everyone neatly fits into predefined categories in society, and the impact on those outside such categories may be obscured due to limited access to these infrastructures. Star's (2007) work on residuals becomes particularly pertinent in this context \cite{star2007enacting}. Residuals, in the context of infrastructure, refer to the often-overlooked consequences or the marginalized experiences that do not neatly align with the mainstream narratives or predefined categories \cite{star2007enacting}. In the case of WhatsApp for Business, the challenges faced by individuals who lack accessibility to constantly adapt to such infrastructures could be marginalized or rendered invisible in the broader discourse. Prior scholarly inquiries have explicated the emergence of multifaceted sociopolitical challenges within internet infrastructures, exacerbating extant inequalities \cite{hogan2018, jasser2023, malazita2019, strubell2020, vonderau2019}. Entities in positions of authority derive advantages from these infrastructural configurations, while those lacking essential resources, knowledge, or representation find themselves inadequately equipped to address ensuing crises and systemic failures arising from such infrastructural developments \cite{jackson2014rethinking, nelson2011body, nelson2002introduction, noble2016future}. As Graham and Marvin's (2001) work explicated infrastructures complex role in upholding and manifesting power structures \cite{graham2001splintering}, changes to the WhatsApp for Business' pipeline has the potential to reduce infrastructural agency for smaller businesses \cite{richardson2016objecting}. 

Meta's revenue plans, for WhatsApp for Business, can raise concerns about exacerbating inequalities, with potential negative effects on smaller businesses while benefiting larger corporations. For instance, while Meta's collaboration with Jio demonstrates feasibility and potential utility, the shifts in their pricing plan pose challenges for similar collaborations with small businesses. Given the high reliance of these businesses on the WhatsApp pipeline, alternative avenues for sustaining and advancing their enterprises become difficult in the face of such pricing adjustments. The question emerges: what ethical obligations do entities, such as Meta bear when introducing and ingraining infrastructures into the daily lives of individuals? Furthermore, the extensive history of product testing in the global south \cite{frugalinnovation} mandates a critical examination from a decolonial perspective, emphasizing the responsibility of companies to ensure that their testing environments are not deliberately harmed subsequent to the integration of their products into mainstream markets. While it is beyond the scope of this study, this expands the ethical inquiry to encompass issues of inclusivity, recognizing that the societal implications of infrastructural integration might disproportionately affect those who are not adequately represented within mainstream frameworks- all of this aggravated by exploitative companies that use the global south to their advantage. Future research should investigate the ethical obligations of entities like Meta to extend beyond their immediate actions, encompassing a responsibility to mitigate the potential invisibility of marginalized experiences and ensure that the benefits and consequences of infrastructural changes are acknowledged, understood, and addressed comprehensively.

%\begin{acks}
%Acknowledgments
%\end{acks}

\bibliographystyle{ACM-Reference-Format}
\bibliography{references.bib}

%\appendix

%\section{Appendix header}

\end{document}